# Regional distribution of the prostaglandin E2 receptor EP1 in the rat brain: accumulation in Purkinje cells of the cerebellum

Eduardo Candelario-Jalil, Helen Slawik, Ingrid Ridelis, Anne Waschbisch, Ravi Shankar Akundi, Michael Hüll, Bernd L. Fiebich *

Department of Psychiatry, University of Freiburg Medical School, Hauptstr. 5, D-79104 Freiburg, Germany

*Corresponding author:

**Bernd L. Fiebich, Ph.D.**
Neurochemistry Research Group
Department of Psychiatry
University of Freiburg Medical School
Hauptstrasse 5
D-79104 Freiburg, Germany
Tel. (49) 761/270-6898
Fax (49) 761/270-6917
E-mail: bernd.fiebich@klinikum.uni-freiburg.de

**Running title:** EP1 receptor expression in the rat brain

**Acknowledgements:** ECJ was supported by a research fellowship from the Alexander von Humboldt Foundation (Germany).

**ABSTRACT**
Prostaglandin $E_2$ (PGE$_2$), is a major prostanoid produced by the activity of cyclooxygenases (COX) in response to various physiological and pathological stimuli. PGE$_2$ exerts its effects by activating four specific E-type prostanoid receptors (EP1, EP2, EP3 and EP4). In the present study, we analyzed the expression of the PGE$_2$ receptor EP1 (mRNA and protein) in different regions of the adult rat brain (hippocampus, hypothalamus, striatum, prefrontal cerebral cortex, parietal cortex, brain stem and cerebellum) using reverse transcription-polymerase chain reaction (RT-PCR), Western blotting and immunohistochemical methods. On a regional basis, levels of EP1 mRNA were the highest in parietal cortex and cerebellum. At the protein level, we found a very strong expression of EP1 in cerebellum as revealed by Western blotting experiments. Furthermore, the present study provides for the first time evidence that the EP1 receptor is highly expressed in the cerebellum, where the Purkinje cells displayed a very high immunolabeling of their perikaryon and dendrites as observed in the immunohistochemical analysis. Results from the present study indicate that the EP1 prostanoid receptor is expressed in specific neuronal populations, which possibly determine the region specific response to PGE$_2$.

**INTRODUCTION**

Prostaglandin endoperoxide H synthases, most commonly known as cyclooxygenases (COX), catalyze the committed step in prostanoid synthesis (for a complete review, see Smith et al., 2000). There are two known COX isoforms, COX-1 and COX-2, which are 90% similar in amino acid sequence and 60% homologous (Smith et al., 2000). COX-1 is expressed constitutively in many organs and contributes to the synthesis of prostanoids involved in normal cellular functions (Seibert et al., 1997), whereas COX-2, which is undetectable in most tissues under normal conditions, can be rapidly induced by pro-inflammatory cytokines *in vitro* or after inflammatory insults *in vivo* (Smith et al., 2000). Interestingly, both COX isoforms appear to be constitutively expressed in normal rat neurons (Yamagata et al., 1993; Breder et al., 1995). COX-2 expression in neurons is rapidly increased by trans-synaptic stimulation (Yamagata et al., 1993; Miettinen et al., 1997), after seizures (Adams et al., 1996), excitotoxicity (Adams et al., 1996; Miettinen et al., 1997; Sanz et al., 1997), spreading depression (Miettinen et al., 1997) and by cerebral ischemia (Collaco-Moraes et al., 1996; Nogawa et al., 1997; Candelario-Jalil et al., 2004), and can rapidly be induced in microglia and astrocyte cultures by a variety of stimuli (Fiebich et al., 1996; Minghetti et al., 1996; Bauer et al., 1997; Yamagata et al., 1993; O'Banion, 1999; Akundi et al., 2005).

Despite the intensive research in this area, the physiological and pathological functions of COX isoforms in the brain are not completely understood, mainly due to the complexity of the system, involving multiple pathways that produce several prostanoids from diverse cell types. In addition, the existence of different prostanoid receptors coupled to different signal transduction pathways adds to the complexity of the role of COX in physiology and/or pathophysiology. A major product of COX activity is prostaglandin $E_2$ ($PGE_2$), a prostanoid that exerts its effects by activating specific receptors, which have been classified into four subtypes (EP1-4).

$PGE_2$ receptors differ in ligand-binding specificity, tissue distribution, and coupling to intracellular signal transduction pathways (Negishi et al., 1995; Sugimoto et al., 2000). All EP receptors are heterotrimeric GTP-binding protein (G-protein)-coupled rhodopsin-type receptors (Watabe et al., 1993; Sugimoto et al., 2000). The primary signaling pathways of the four EP receptors are as follows: EP1 couples to elevation of intracellular calcium levels $[Ca_{2+}]_i$ (Funk et al., 1993), EP2 and EP4 couple to an increase in intracellular cAMP accumulation (Regan et al., 1994; Bastien et al., 1994), and EP3 couples to a decrease in intracellular cAMP concentrations (Boie et al., 1997).

Up to now, EP receptors have been found on a cellular level only in neurons throughout the unlesioned rodent CNS. In contrast to EP2 and EP4 receptors, which have been found only in restricted neuronal cell groups (Zhang and Rivest, 1999), EP3 receptors show a widespread constitutive expression in neurons throughout the brain (Nakamura et al., 2000; Ek et al., 2000). Unlike the other EP receptors, there is very little, if any, existing documentation on the regional distribution of EP1 in the rat brain. The purpose of the present study was therefore to identify the rat brain areas that express EP1 under basal conditions, and to compare the expression levels of this prostanoid receptor (EP1 mRNA and protein) in seven discrete areas of the adult rat brain, using reverse transcription-PCR, western blotting and immunohistochemistry. An analysis of the distribution of EP1 receptor in the brain is essential to assess its functional role in physiological and pathophysiological conditions.

# MATERIALS AND METHODS

## Dissection of rat brain regions

Sprague-Dawley rats of either sex were sacrificed by cervical dislocation and seven different brain regions (hippocampus, hypothalamus, striatum, prefrontal cortex, parietal cortex, brain stem and cerebellum) were dissected out according to the dissection method of Glowinski and Iversen (Glowinski and Iversen, 1966) as reported before in our previous studies (Candelario-Jalil et al., 2001a, b). Tissue samples were placed immediately in liquid $N_2$ and later kept at -80°C until further analysis.

## RNA extraction and RT-PCR analysis

Total RNA from brain regions was extracted using the guanidine isothiocyanate method according to Chomczynski and Sacchi (Chomczynski and Sacchi, 1987). For RT-PCR, 2 µg of total RNA was reverse transcribed using M-MLV reverse transcriptase (Promega, Mannheim, Germany) and random hexamers (Promega). PCR was carried out using Taq polymerase (Promega), dNTP master mix (Invitek, Berlin, Germany). Primers were designed using PrimerSelect Software from DNA Star Inc. (Madison, WI, USA). The following specific primers for rat EP1 receptor were used: forward: 5'-CTG GGC GGC TGC ATG GTC TTC TTT-3', reverse: 5'-GCG GAG GGC AGC TGT GGT TGA-3', 65°C, 40 cycles, product length: 497 bp. Equal equilibration was determined using rat $\beta$-actin primers (forward, 5'- ATG GAT GAC GAT ATC GCT-3'; reverse, 5'-ATG AGG TAG TCT GTC AGG T-3', 48°C , 30 cycles, amplicon size 569 bp). PCR products were separated electrophoretically on a 2% agarose gel. Potential contamination by genomic DNA was controlled by omitting reverse transcription and using $\beta$-actin primers in the subsequent PCR amplification. Only RNA samples showing no bands after this procedure were used for further investigation.

## Western Blot analysis

Brain regions were dissected out as described above and were diluted 1:10 (wt/v) with SDS sample buffer (42 mM Tris-HCl pH 6.8, 1.3% sodium dodecyl sulfate buffer, 100 µM orthovanadate, 6.5% Glycerine) (Laemmli, 1970) and homogenized with a sonicator (Branson Sonifier 250 at the 50% pulse mode with a microtip; Branson Inc., Danbury, CT). Tissue homogenates were centrifuged at 14 000 rpm at 4°C for 10 min and supernatants were taken for the analysis. Protein content was determined using the bicinchoninic acid method (BCA protein assay kit, Pierce, distributed by KFC Chemikalien, München, Germany). Before loading, samples were incubated at 95°C for 5 min. 60 µg of protein were subjected to SDS-PAGE on a 10 % gel under reducing conditions. Proteins were transferred onto a polyvinylidene fluoride membrane (Millipore, Bedford, MA, USA) by semi-dry blotting. The membrane was blocked overnight at 4°C using Rotiblock (Roth, Karlsruhe, Germany). Rabbit anti-EP1 receptor polyclonal antibody (Cayman, Ann Arbor, MI, USA) was diluted (1:500) in TBS-Tween 20 containing 1% BSA. Membranes were incubated for 2 h at room temperature with the primary antibody, washed three times in TBS-Tween 20 and for 1 h with the secondary antibody HRP-linked anti-rabbit IgG (Amersham Biosciences, Freiburg, Germany). Subsequent detection was performed using the ECL Western blotting system (Amersham Biosciences) according to the manufacturer's instruction. Specificity of the antibody was assessed by omitting the first antibody in Western blotting experiments. All Western blot experiments were carried out in brain areas from three different rats. Quantification of the Western Blots was performed using ScanPack 3.0 software (Biometra, Göttingen, Germany) and for descriptive purposes means and standard deviations were calculated.

**EP1 Immunohistochemical Analysis**

Animals were deeply anesthetized with chloral hydrate (300 mg/kg; i.p.) and were intracardially perfused with cold PBS followed by 4% (w/v) paraformaldehyde (PFA) in PBS. Brains were removed and immersed in 4% PFA-PBS for another 2 days at 4°C. After cryoprotection in 20% sucrose for at least two days, the brains were rapidly frozen in isopentane and stored at -80°C. Sagittal cryostat sections (10 μm) were cut, mounted on Superfrost sections and processed for immunohistochemistry with the avidin-biotinperoxidase method according to the manufacturer's instructions as described before (Slawik et al., 2004). First the sections were reacted in 1% $H_2O_2$ to quench endogenous peroxidase activity for 20 min. After washing, the non-specific binding sites were blocked by incubation with 4% respective normal serum in PBS for 30 min. Then sections were incubated overnight at 4°C with the primary antibody (rabbit polyclonal antibody against EP1; Cayman Chemical, diluted 1:300) in 1% bovine serum albumin (BSA) in PBS. After incubation with the primary antibody, the sections were washed three times and incubated with the secondary antibody (goat anti-rabbit, Santa Cruz Biotechnology, diluted 1:200) for one hour. After three washes the sections were incubated for 30 min with the avidin-biotinylated horseradish peroxidase complex (ABC-Elite kit, Vector Laboratories, Burlingame, CA, USA). All washes between antibody incubations were made in PBS. For visualization of peroxidase, all sections were incubated with 0.05% diaminobenzidine and 0.02% $H_2O_2$. Counterstaining was performed with Mayer's hematoxylin. After dehydration the sections were coverslipped with Vitro-Clud mounting media (Emmendingen, Germany). Negative controls, consisting of sections incubated in the absence of primary antibodies, gave no signal and thus confirmed the specificity of the here shown antigen-antibody-binding signal.

**RESULTS AND DISCUSSION**

Using semi-quantitative RT-PCR, we observed the presence of single transcripts for EP1 (497 bp) and β-actin (569 bp) migrating at the predicted position (Fig. 1). EP1 mRNA was detected in all brain samples examined although there were marked differences in the level of expression of this prostanoid receptor among the cerebral regions studied. On a regional basis, the strongest levels of EP1 mRNA were found in parietal cortex and cerebellum, followed in descending order by frontal cortex and striatum (Fig. 1). The hypothalamus, hippocampus and brain stem displayed a low-level EP1 mRNA signal as depicted in Fig. 1.

In order to confirm that the pattern of EP1 mRNA expression is also observed at the protein level, we performed Western blotting of lysates prepared from different brain areas. The result of a representative Western blotting experiment is shown in Fig. 2. In line with the data observed at the mRNA level, we found pronounced cerebellar levels of the EP1 protein, which was consistent in all animals studied. Although EP1 protein was also observed in parietal cortex, it did not correspond to the strong mRNA signal in this area (Fig. 2), which was comparable to cerebellum (Fig. 1). On the other hand, our results showed a very weak EP1 protein expression in the other brain regions under study (Fig. 2), which fit to the mRNA findings (Fig. 1).

In addition, we performed an immunohistochemical analysis to further confirm the results of the Western blotting on a cellular level. Few EP1-positive cells were observed in parietal cortex and very few immunopositive cells were found in the other regions studied (data not shown). In contrast, and in line with our results from mRNA and western blot studies, EP1 receptor immunoreactivity was markedly found in the cerebellum, where the Purkinje cells displayed a very high immunolabeling of their perikaryon and dendrites as observed in Fig. 3. No other cell types in the cerebellum showed immunostaining for EP1.

To the best of our knowledge, this is the first study to consistently characterize the expression of the EP1 receptor in different areas of the rat brain. It is important to mention that compared to all of the other brain regions examined, levels of EP1 protein were the highest in the cerebellum. The physiological function of the increased EP1 levels in cerebellum is unclear at this time.

One of the hallmarks of cerebellar Purkinje cells is their ability to express a characteristic form of activity-dependent synaptic plasticity named long-term depression (LTD) which is essential for motor learning. The cerebellum is the brain region where learned movements are stored and LTD is a key mechanism involved in this function (Saab and Willis, 2003). Synaptically-induced dendritic $Ca_{2+}$ signaling has been known for a long time to be a critical step in LTD induction (Sakurai, 1990; Konnerth et al., 1992). Several experimental evidences indicate that not merely the lack of synaptically mediated $Ca_{2+}$ signaling in Purkinje cells, but also a change in the temporal dynamics of postsynaptic $Ca_{2+}$ transients after genetic deletion of the calcium-binding protein calbindin D-28k (calbindin), might cause impairment of motor coordination (Airaksinen et al., 1997; Barski et al., 2003).

$PGE_2$ stimulates $Ca_{2+}$ mobilization via EP1 receptor (Watabe et al., 1993; Katoh et al., 1995), but the mechanism of the increase in intracellular $Ca_{2+}$ and the identification of a G-protein which interacts with the EP1 receptor remains to be elucidated. EP1-induced $Ca_{2+}$ influx correlates with only a small increase in $IP_3$ generation and is dependent on the presence of extracellular $Ca_{2+}$. It is thus believed that the EP1 receptor couples to $Ca_{2+}$ through a mechanism independent of $G_q$ protein (Narumiya et al., 1999).

Since EP1 activation is linked to an increase of intracellular $Ca_{2+}$ levels, we speculate that $PGE_2$ might be involved in the modulation of neuronal plasticity of Purkinje cells through activation of EP1 receptors. The EP1 receptor appears to be the most likely candidate due to its high expression in these cells as shown here (Fig. 3) and due to the very low or null expression of the other EP receptors (EP2-4) in cerebellum (Ek et al., 2000; Zhang and Rivest, 1999).

The notion that $PGE_2$ could modulate synaptic plasticity in cerebellum is based on previous studies on the role of this prostanoid in long-term potentiation (LTP) in hippocampus (another example of synaptic plasticity and of the putative biological processes underlying memory). It has been previously shown that COX-2, but not COX-1, mediates $PGE_2$ signaling in hippocampal long-term synaptic plasticity (Chen et al., 2002). Selective inhibitors of COX-2 significantly reduced membrane excitability, back-propagating dendritic action potential-associated $Ca_{2+}$ influx, and LTP induction in hippocampus, while a COX-1 inhibitor was ineffective. All these effects were effectively reversed by exogenous application of $PGE_2$, but not by other prostaglandins (Chen et al., 2002). Similar results were found by Shaw et al. (Shaw et al., 2003), suggesting that $PGE_2$ plays an important regulatory role in synaptic plasticity.

Since $PGE_2$ is not stored or secreted from synaptic vesicles like the classic neurotransmitters, and rapidly diffuses and activates EP receptors, some authors have suggested that $PGE_2$ signaling must depend on basal COX-2 expression. Interestingly, COX-2 mRNA and protein are normally expressed in relatively high levels in several neuronal populations throughout the CNS (Yamagata et al., 1993; Breder et al., 1995) and COX-2 is consistently localized in dendritic spines of neurons that receive synaptic input including Purkinje cells (Kaufmann et al., 1996; Pardue et al., 2003).

$PGE_2$ is also involved in the regulation of membrane excitability in sensory neurons (Gold et al., 1998; Nicol et al., 1997). Furthermore, $PGE_2$ through its action on EP1 receptors, induces $Ca_{2+}$ release from intracellular ryanodine/caffeine-sensitive stores and stimulates catecholamine release from intracellular ryanodine/caffeine-sensitive stores and stimulates catecholamine release in adrenal medullary cells (Shibuya et al., 1999; Negishi et al., 1990). Some of the detrimental effects of $PGE_2$ have been associated to its ability to bind predominantly to the EP1 receptors. These include $PGE_2$-mediated fever (Oka and Hori, 1994; Batshake et al., 1995), allodynia (Minami et al., 2001), and acute inflammatory pain (Stock et al., 2001).

Results from very recent studies indicate that activation of EP1 receptors contributes to COX-2-dependent neurotoxicity following excitotoxic neuronal injury *in vitro* (Carlson, 2003; Zhou et al., 2004) and *in vivo* (Kawano et al., 2004). The mechanisms by which EP1 receptors promote neuronal injury are likely to include amplification of the intracellular $Ca_{2+}$ levels induced by excitotoxicity (Kawano et al., 2004).

The present study provides evidence that the EP1 receptor is highly expressed in Purkinje cells of the rat cerebellum, which could inspire new investigations to elucidate the specific role of $PGE_2$/EP1 signaling in these cells. A deeper understanding of Purkinje cell physiology, including a delineation of the whole spectrum of signal transduction pathways, is essential in order to understand the organization of cerebellar circuits. As demonstrated here, the other rat brain areas examined only showed a low EP1 expression under normal conditions (Figs. 1 and 2). Only the parietal cortex displayed a very strong mRNA signal (compared to the one observed in cerebellum), but results from the western blotting and the immunohistochemical analysis showed that EP1 protein expression is relatively low as compared to cerebellum. This suggests that the expression of the EP1 receptor might not only be regulated at the transcriptional level. Some factors such as region-specific changes in the translation of the mRNA, post-translational modification and/or processing of the EP1 protein into functional membrane receptors might also be involved and could explain this observation.

In summary, results from the present study indicate that the EP1 prostanoid receptor is expressed in specific neuronal populations, which possibly determine the region specific response to $PGE_2$. The present findings lay the foundation for future studies aimed to explain the cellular and molecular mechanisms by which $PGE_2$, through its interaction with EP1 receptors, could modulate neuronal function.

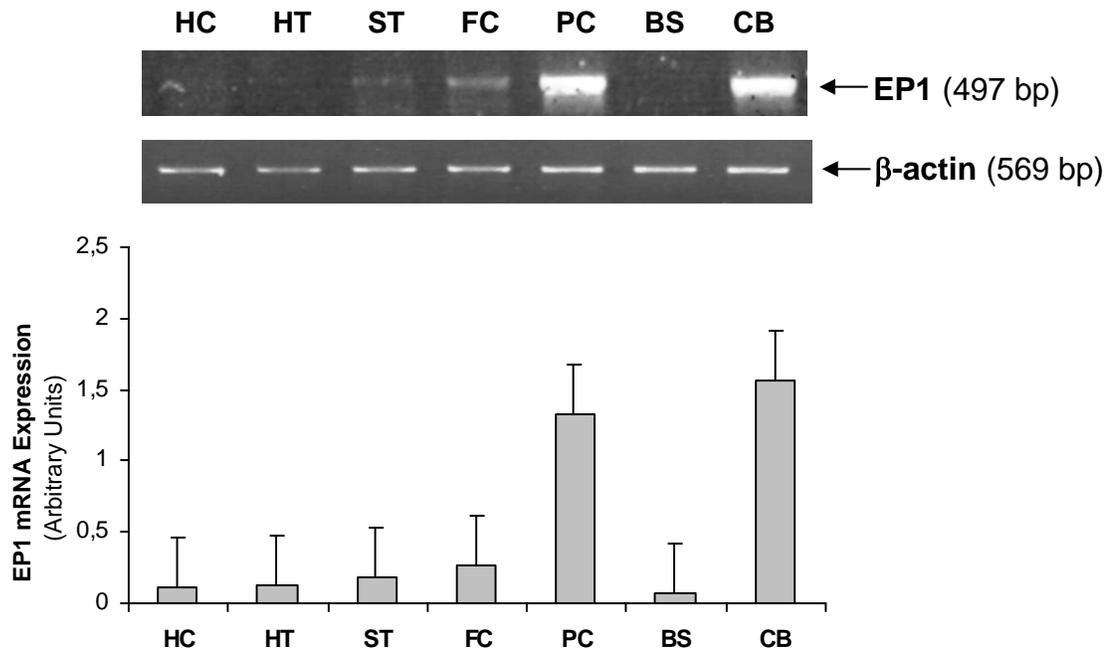

**Fig. 1.** EP1 mRNA expression in the rat brain. (**A**) Representative gel electrophoresis of RT-PCR products for EP1 and b-actin. The expected size of the amplicon was 497 bp for EP1 and 569 bp for b-actin. (**B**) Densitometric analysis of the EP1 RT-PCR data. To control for possible differences in initial amount of total RNA template used for RT-PCR in each sample used, values were normalized against b-actin. Values are mean ± S.D. from 3 different animals. HC: hippocampus; HT: hypothalamus; ST: striatum; FC: prefrontal cortex; PC: parietal cortex; BS: brain stem; CB: cerebellum. **p<0.01 with respect to HC, HT, ST, FC and BS.

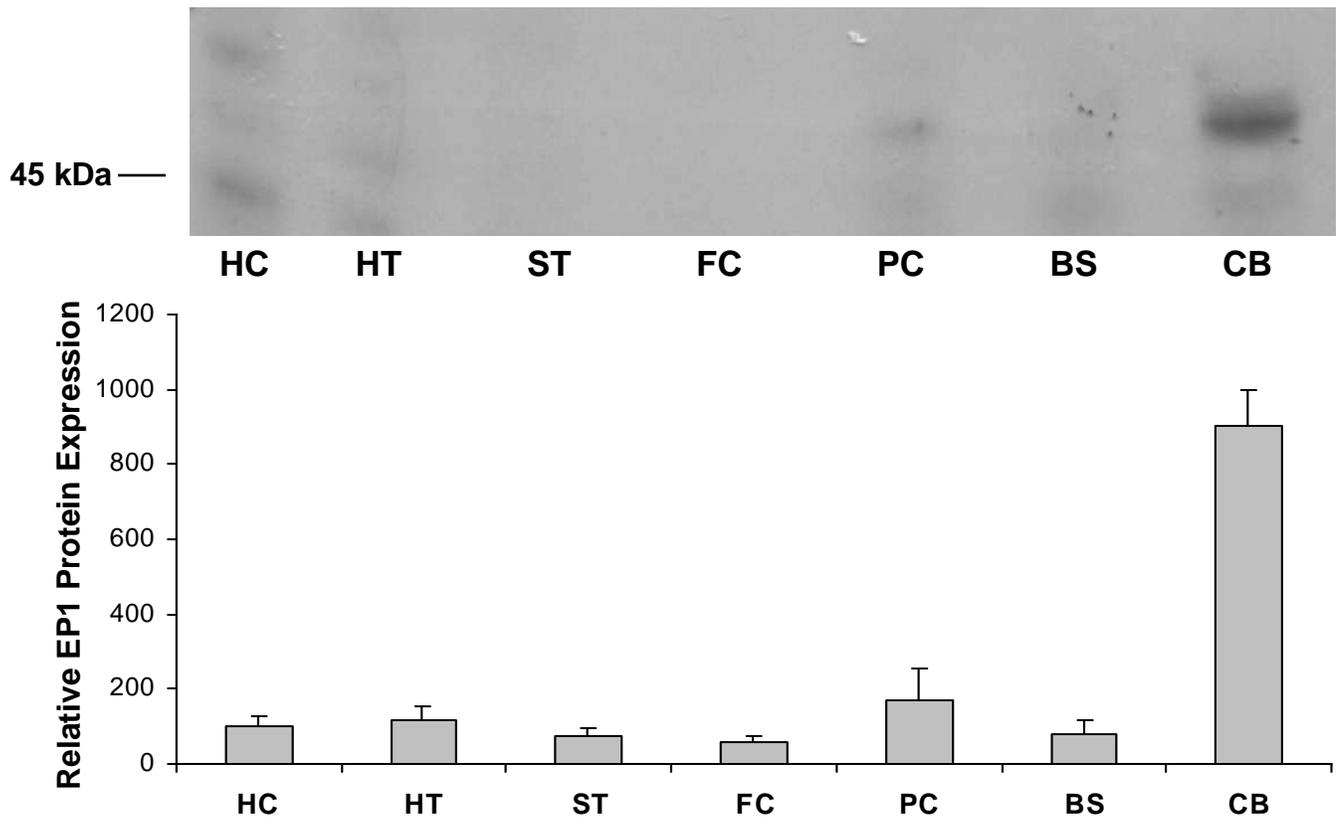

**Fig. 2.** EP1 protein levels in different rat brain regions. (**A**) Proteins extracted from homogenised brain tissue were subjected to western blot analysis. An EP1 immunoreactive band with an apparent molecular weight of approximately 42 kDa was detected. (**B**) Densitometric analysis of the EP1 receptor protein content in the rat brain. Values are mean ± S.D. from 3 different animals. HC: hippocampus; HT: hypothalamus; ST: striatum; FC: prefrontal cortex; PC: parietal cortex; BS: brain stem; CB: cerebellum. **p<0.01 with respect to the other brain regions.

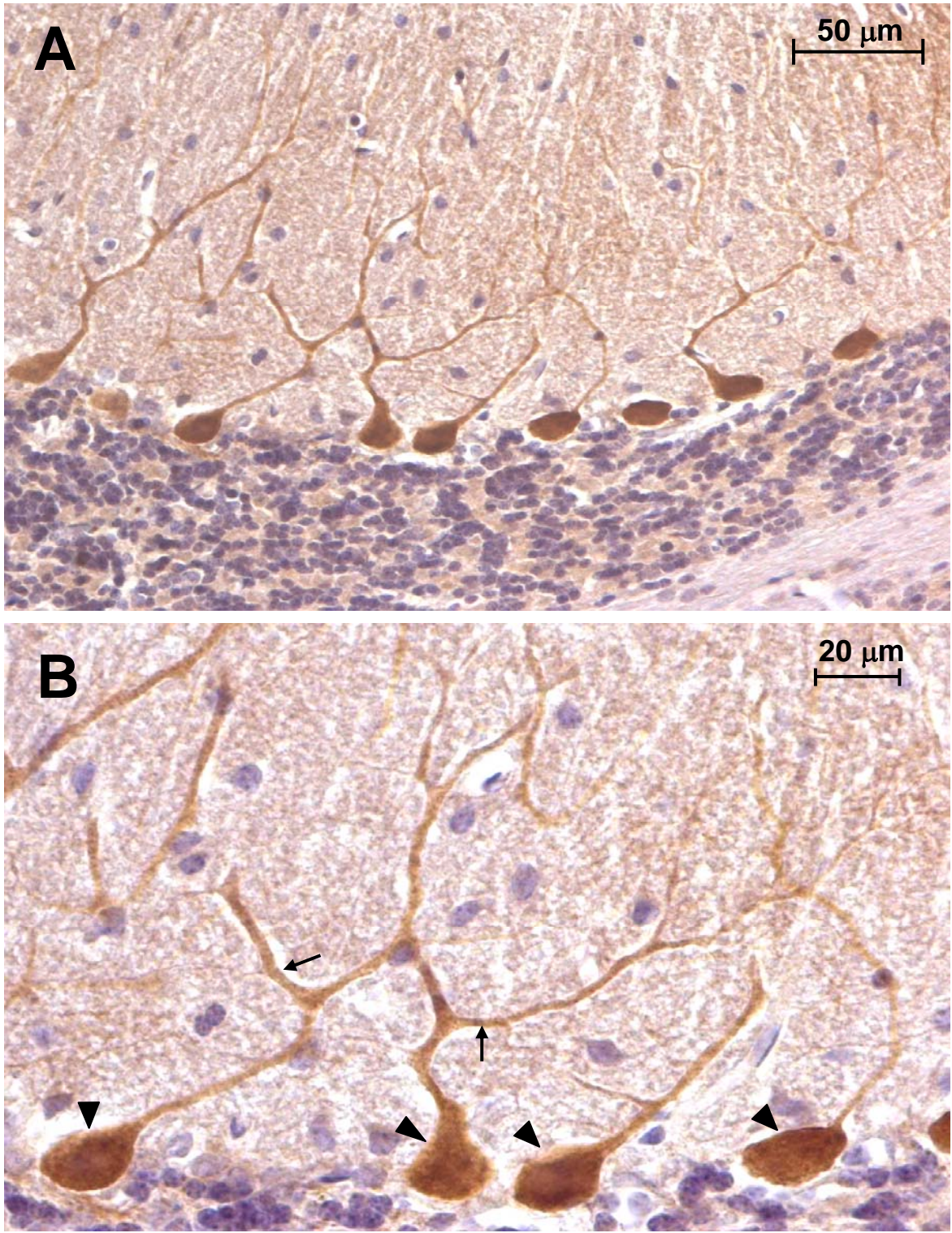

**Fig. 3.** Strong immunoreactivity to the EP1 receptor is detected in Purkinje cells of the rat cerebellum (**A**). EP1 immunoreaction (brown) in Purkinje cells is observed in both perikaryon(arrowheads) and dendrites (arrows) (**B**).